\documentclass[reprint,amsmath,amssymb,aps,pra,nofootinbib,nobibnotes,floatfix]{revtex4-2}
\usepackage[T1]{fontenc}
\usepackage[utf8]{inputenc}
\usepackage{textcomp}
\usepackage{fix-cm}
\usepackage{graphicx}
\usepackage{dcolumn}
\usepackage{bm}
\usepackage{siunitx}
\usepackage[version=3]{mhchem}
\usepackage{booktabs,multirow,longtable}
\usepackage{xcolor}
\usepackage{setspace}
\usepackage[unicode=true,hidelinks]{hyperref}
\usepackage{etoolbox}
\makeatletter
\AtBeginDocument{%
  \let\auto@bib@innerbib\@empty
  \let\auto@bib\@empty
  \let\write@bibliographystyle\relax

}

\makeatother

\setcitestyle{super,sort&compress}
\begin{document}
\title{Protonic thermoelectric effect of Superionic H\textsubscript{2}O and magnetic field generation in Uranus and Neptune}

\author{Daohong Liu\textsuperscript{1,2}}

\author{Wei Zhang\textsuperscript{3}}

\author{Yu He\textsuperscript{1,4,5*}}

\author{Xinzhuan Guo\textsuperscript{1,4}}

\author{Chuanyu Zhang\textsuperscript{6}}

\author{Yang Sun\textsuperscript{7}}

\affiliation{\textsuperscript{1}State Key Laboratory of Critical Mineral Research and Exploration, Institute of Geochemistry, Chinese Academy of Sciences, Guiyang 550081, China.\\\textsuperscript{2}College of Earth and Planetary Sciences, University of Chinese Academy of Sciences, Beijing 100049, China.\\\textsuperscript{3}School of Geography and Environmental Science (School of Karst Science), Guizhou Normal University, Guiyang 550025, China.\\\textsuperscript{4}Key Laboratory of High-Temperature and High-Pressure Study of the Earth's Interior, Institute of Geochemistry, Chinese Academy of Sciences, Guiyang 550081, Guizhou, China.\\\textsuperscript{5}Center for High Pressure Science and Technology Advanced Research, Shanghai 201203, China.\\\textsuperscript{6}College of Physics, Chengdu University of Technology, Chengdu, 610059, China.\\\textsuperscript{7}School of Materials, Sun Yat-sen University, Shenzhen, China.}

\begin{abstract}
The origin of the anomalous magnetic fields of Uranus and Neptune, with significant tilts and multi-dipole configurations, is central to understanding the internal structure and evolution of the ice giants. Recent investigations confirmed that superionic H\textsubscript{2}O ice is thermodynamically stable and constitutes the dominant H\textsubscript{2}O phase within their icy mantles. Here, we use deep learning and~\emph{ab initio}~molecular dynamics to investigate the thermal and protonic transport properties of superionic H\textsubscript{2}O ice. We demonstrate that superionic H\textsubscript{2}O ice exhibits a pronounced protonic thermoelectric effect, with a maximum Seebeck coefficient that can reach \ensuremath{\sim}620 \textmu V·K\textsuperscript{-1} within the interior of Uranus, whereas Neptune's is lower, within the range of 570--585 \textmu V·K\textsuperscript{-1}. Consequently, the temperature gradients in the icy mantles can induce proton migration, thereby driving magnetic field generation. Based on this mechanism, and within a dipolar approximation that captures the mean field, the disparities in mean dipolar magnetic field strength between Uranus and Neptune can be explained exclusively by their differing internal temperature gradients, and the predicted mean field strengths are comparable to those observed by Voyager 2.
\end{abstract}

\maketitle

\begin{center}\small *Corresponding author's Email: \href{mailto:heyu@mail.gyig.ac.cn}{\nolinkurl{heyu@mail.gyig.ac.cn}}\end{center}

\section*{Introduction}

Unlike the predominantly dipolar and nearly axisymmetric magnetic fields of terrestrial planets and gas giants\cite{main001}, the magnetic fields of Uranus and Neptune exhibit significant non-dipole and non-axisymmetric characteristics, with their magnetic axes substantially offset from the planetary center and highly inclined relative to the rotation axis\cite{main002,main003,main004,main005,main006,main007}. This unique magnetic field configuration not only challenges conventional dynamo theory, but also offers critical insights into the internal dynamics, material states, and evolutionary history of the ice giants. To explain the anomalous geometry of magnetic fields of Uranus and Neptune, Stanley and Bloxham\cite{main008} proposed the thin-shell dynamo model. It posits that the magnetic field is generated within a relatively thin, convecting fluid layer in the outer part of the icy mantles, beneath which lies a large, stably stratified region. This thin-shell geometry produces smaller-scale convection, thereby generating a multipolar magnetic field. For the thin-shell dynamo model, the formation of stable stratification beneath the convective layer and the mechanical and electrical properties of the underlying stratified region are also key factors for the stable generation of magnetic fields\cite{main005,main009}. Alternatively, the magnetic field may be generated by internal turbulence that is not strongly constrained by the planet's rotation\cite{main010}, thus leading to a multipolar field. Nevertheless, the lack of knowledge of the internal composition and transport properties of ice giants continues to impede our ability to discriminate among the magnetic field generation mechanisms and to identify the physical origin of the observed magnetic fields.

Interior structure models and high-pressure studies indicate that H\textsubscript{2}O, CH\textsubscript{4}, and NH\textsubscript{3} ~ices are the dominant components in the interior of Uranus and Neptune\cite{main011,main012,main013,main014}. Besides, recent high-pressure experiments and theoretical calculations have confirmed the superionic transition of H\textsubscript{2}O and NH\textsubscript{3}  under conditions relevant to the interiors of Uranus and Neptune\cite{main012,main013,main015,main016,main017,main018,main019,main020,main021,main022,main023}. In these superionic ices, protons diffuse like a liquid through solid-like sublattices of heavier nuclei. As reported previously, the solid-like sublattice retains stability under applied shear stress, which yields characteristic solid-like mechanical properties\cite{main024,main025}. In particular, superionic H\textsubscript{2}O ice undergoes a phase transition from a body-centered cubic (bcc) to a face-centered cubic (fcc) phase at \ensuremath{\sim}2000 K with increasing temperature\cite{main016,main018,main020,main022}. This phase transition is accompanied by a significant increase in protonic conductivity\cite{main018,main026}. The entropy contributed by the highly diffusive protons leads to high stability of the superionic fcc phase, which remains stable across most of the pressure-temperature (\emph{P--T}) range of the icy mantles\cite{main016,main017,main018,main019}. These findings imply that the superionic fcc H\textsubscript{2}O ice is likely to be a major component of the icy mantles beneath the fluid H\textsubscript{2}O layer. Therefore, its electrical and mechanical properties are proposed to influence the magnetic field generation in the ice giants\cite{main005,main024,main025,main026,main027,main028,main029}.

The ionic thermoelectric (\emph{i}-TE) effect refers to a thermoelectric effect based on ionic transport\cite{main030,main031,main032,main033}, specifically a physical phenomenon wherein charged ions in ionic conductors (such as solid electrolytes, gels, ionic liquids, etc.) undergo directed migration driven by a definite temperature gradient, thereby establishing a thermovoltage across the material\cite{main031,main032,main033}. The~\emph{i}-TE materials, which present high Seebeck coefficients rivaling tens of mV·K\textsuperscript{-1}, have been used for electricity generation devices\cite{main030,main031,main034}. Since superionic H\textsubscript{2}O ice is an effective proton conductor in the ice giants, the protonic thermoelectric (\emph{p}-TE) mechanism might generate planetary-scale thermocurrents wherever a persistent thermal gradient exists. Such thermocurrents would not merely be a passive byproduct of interior dynamics but also influence the energy source and topology of magnetic fields.

In this study, we calculated the protonic conductivity, thermal transport properties, and~\emph{p}-TE Seebeck coefficients of superionic H\textsubscript{2}O ice under the conditions of the ice giants' interior using combined \emph{ab initio} molecular dynamics (AIMD) and deep potential molecular dynamics (DPMD) methods. In addition, based on a symmetric thermoelectric current model of a spherical superionic ice shell with a radial temperature gradient and radially varying transport coefficients, we also evaluated the potential impact of the \emph{p}-TE effect on planetary-scale magnetic field generation in the icy mantles of Uranus and Neptune. This mechanism may couple with above the thin-shell conductive fluid dynamo\cite{main008}, yielding insights into the complex magnetic field generation mechanisms operating inside ice giants.

\section*{Results and discussion}

\subsection*{Density Profiles of Superionic H\textsubscript{2}O Ice}

We first evaluated our trained potential and computational method by comparing the pressure--temperature--density (\emph{P--T}--\emph{\ensuremath{\rho}}) relations of bcc and fcc H\textsubscript{2}O calculated using AIMD and DPMD simulations at 150--450 GPa and 1000--4000 K, as shown in Fig. 1a. The density profiles of bcc and fcc H\textsubscript{2}O calculated using DPMD show great agreement with AIMD results. The accuracy of the deep potential (DP) model was first assessed with~\emph{ab initio}~validation datasets (Supplementary Figs. S1 and S2) and was further evaluated by comparing the predictions of radial distribution functions (RDFs) with AIMD results (Supplementary Figs. S3--S8). The calculated densities are also consistent with experimental results using static and dynamic compression methods\cite{main016,main020}. In addition, we found that fcc H\textsubscript{2}O becomes denser than bcc H\textsubscript{2}O at pressures above \ensuremath{\sim}200 GPa (Fig. 1a), which indicates that bcc H\textsubscript{2}O exhibits lower compressibility than fcc H\textsubscript{2}O as pressure increases. These results are consistent with previous experiments\cite{main015,main016,main020}.

On the other hand, recent experimental studies also emphasized the importance of volume expansion for the superionic transition of both bcc and fcc H\textsubscript{2}O\cite{main022}. In this case, we determined the volume changes for bcc and fcc H\textsubscript{2}O upon the superionic transition at 200 GPa with refined temperature intervals (Fig. 1b). Indeed, the onset of the superionic state is accompanied by volume expansion. At \ensuremath{\sim}200 GPa, the onset of superionic transition in bcc ice occurs at 2250--2600 K, approximately 500 K higher than the transition in the fcc phase, indicating that the superionic bcc phase is thermodynamically metastable under these conditions, consistent with experimental observation\cite{main020,main022}. The thermal expansion coefficients~\emph{\ensuremath{\alpha}}~(Fig. 1c), obtained from the fitted curves, exhibit a characteristic~\emph{\ensuremath{\lambda}}-shaped peak associated with the onset of the superionic transition. The calculated peak value of~\emph{\ensuremath{\alpha}}~in this study is within the range of previous predictions. Our results have confirmed that the volume change upon the superionic transition is critical for the determination of the superionic transition temperature for experimental observations.

\begin{figure*}[tp]
\centering
\includegraphics[width=0.85\textwidth]{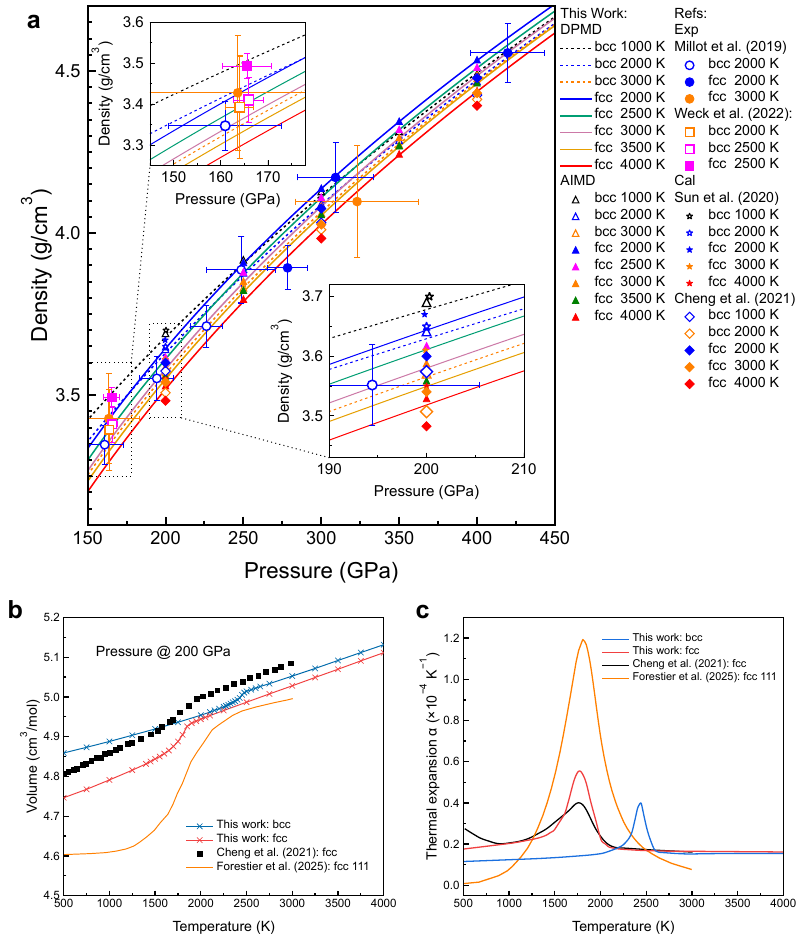}
\caption{\textbf{Thermodynamic properties of bcc and fcc H\textsubscript{2}O at 150--450 GPa and 2000--4000 K. a,} Density--pressure relationships of the bcc and fcc H\textsubscript{2}O at high \emph{P}--\emph{T}. The colors of the curves and symbols correspond to temperature along isotherms. Dashed and solid curves represent the DPMD calculation results for the bcc and fcc phases, respectively. Open and solid triangle symbols represent the AIMD calculation results. The previous experimental results are noted with circles and squares\cite{main016,main020}. The previous computational results are noted with symbols of star and diamond\cite{main018,main024}. \textbf{b,} Calculated molar volumes of bcc (blue cross-dotted line) and fcc (red cross-dotted line) are presented as a function of temperature along isobar at 200 GPa. A sharpening of the change in molar volume with increasing temperature suggests a superionic transition. \textbf{c,} The volumetric thermal expansion coefficients are represented by blue and red curves for bcc and fcc H\textsubscript{2}O in comparison with previous results\cite{main018,main022}.}
\label{fig:main1}
\end{figure*}

\subsection*{Protonic Thermoelectricity in Superionic H\textsubscript{2}O Ice}

Phase boundaries for bcc--fcc and fcc--liquid phase transitions are determined using non-equilibrium thermodynamic integration (NETI) and Clausius--Clapeyron equation (CCE) methods (see Supplementary information for details) at 100--500 GPa as shown in Fig. 2. Our results are consistent with previous experimental\cite{main016,main020} and computational\cite{main018,main023,main035} results (Supplementary Figs. S9--S13). Based on the phase boundaries, we calculated the thermal conductivity of bcc and fcc H\textsubscript{2}O by employing the non-equilibrium molecular dynamics (NEMD) method (Supplementary Figs. S14--S23). The reliability of the NEMD simulations based on the DP model was validated in Supplementary Fig. S24.

In bcc H\textsubscript{2}O, the simulated results of thermal conductivity exhibit an obvious decrease with increasing temperature (Fig. 2 and Supplementary Fig. S23 and Table. S1).~This behavior arises from enhanced anharmonic scattering at high temperatures. In contrast, for superionic fcc H\textsubscript{2}O, the results of thermal conductivity exhibit a different behavior, which show a near temperature independence of the thermal conductivity (Fig. 2 and Supplementary Fig. S23 and Table. S2). The high proton mobility may render proton--phonon scattering the dominant mechanism suppressing thermal transport\cite{main036}. This scattering effect becomes intrinsically stronger with increases in proton diffusion coefficients. Consequently, the lattice thermal conductivity of the fcc H\textsubscript{2}O exhibits insignificant temperature dependence, as the proton diffusion coefficients vary only slightly with increasing temperature. In addition, pressure has a monotonic enhancing effect on the thermal conductivity for both bcc and fcc H\textsubscript{2}O (Supplementary Fig. S18). As pressure rises, decreasing interatomic distances constrain atomic vibrational amplitudes. This effect may suppress anharmonic phonon scattering, which ultimately leads to enhanced lattice thermal conductivity in H\textsubscript{2}O ices. Based on the thermal profiles of Uranus and Neptune\cite{main037}, H\textsubscript{2}O mainly exists in the form of superionic fcc H\textsubscript{2}O in icy mantles. The highly diffusive protons in fcc H\textsubscript{2}O lead to heat flux dissipation, thus decreasing the thermal conductivity efficiency in the interior of the ice giants.

\begin{figure*}[tp]
\centering
\includegraphics[width=0.81\textwidth]{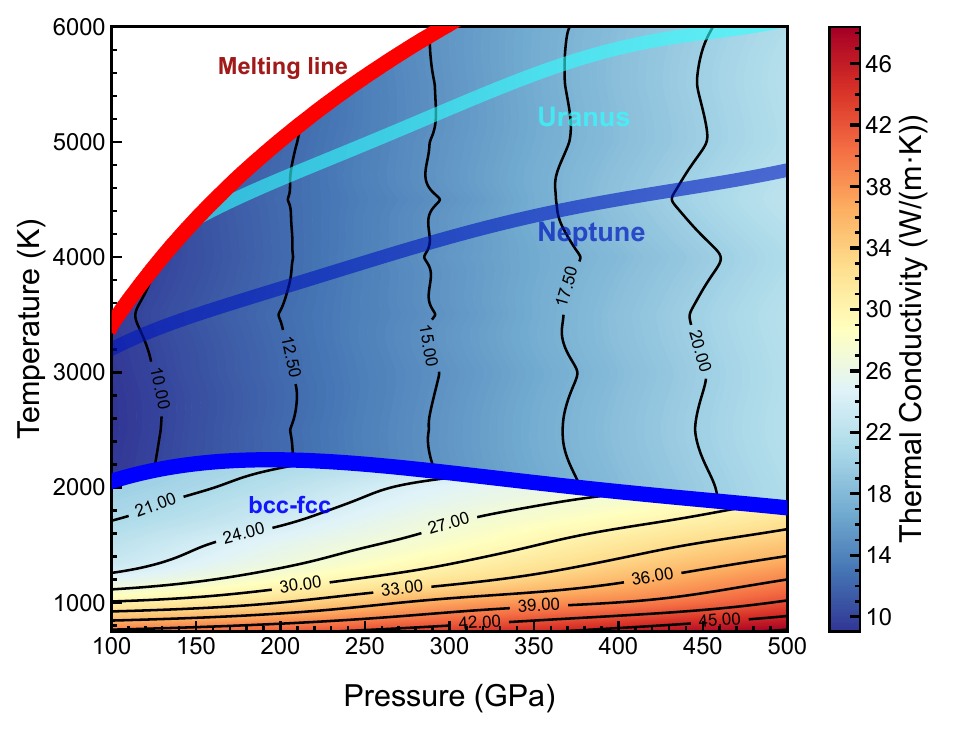}
\caption{\textbf{Thermal conductivity of bcc and fcc H\textsubscript{2}O under conditions relevant to the interiors of Uranus and Neptune.} The contour lines indicate the magnitude of thermal conductivity. The blue and red curves exhibit the bcc-fcc and fcc-liquid phase boundaries, respectively. The cyan and dark blue shaded bands represent the thermal profiles of Uranus and Neptune, respectively, which are averaged from the U1 and U2, and N1 and N2b interior thermal profile models reported in Ref.~\citenum{main037}.}
\label{fig:main2}
\end{figure*}

We calculated proton diffusion coefficients in H\textsubscript{2}O by fitting the mean square displacements (MSDs) of protons under the internal \emph{P--T} conditions relevant to Uranus and Neptune (Fig. 3a and Supplementary Figs. S25--S27). The protonic conductivity of superionic fcc H\textsubscript{2}O ice is estimated using the Nernst-Einstein relation, and the resulting estimates are fitted using the Arrhenius equation (Fig. 3b and Supplementary Table S3). Despite a reduction in the proton diffusion coefficient upon compression (Fig. 3a), the increase in proton concentration is sufficiently substantial to offset the decline in proton diffusivity, which consequently improves the overall protonic conductivity (Fig. 3a inset).

Furthermore, to elucidate the thermoelectric response of superionic fcc H\textsubscript{2}O, the Widom insertion method has been employed to calculate the proton chemical potential, enabling us to determine the partial molar entropy of protons (Fig. 3c). Based on these values, we calculated the Seebeck coefficient of the superionic fcc H\textsubscript{2}O within the framework of Onsager relations\cite{main030} under the interior conditions of Uranus and Neptune (Fig. 3d). The superionic fcc H\textsubscript{2}O maintains a significant and relatively pressure--insensitive Seebeck response throughout the range of planetary interior conditions, underscoring the persistence of a significant thermoelectric effect even in the deep interior of icy mantles. In the interior of ice giants, the Seebeck coefficient of fcc H\textsubscript{2}O slightly increases with depth. The maximum Seebeck coefficient in Uranus can reach 620 \textmu V·K\textsuperscript{-1}, while in Neptune, the coefficient is lower, ranging between 570--585 \textmu V·K\textsuperscript{-1}.

Importantly, the high Seebeck coefficient is complemented by high protonic conductivity, resulting in a large thermoelectric power factor (Supplementary Fig. S28). Quantitative comparison reveals that the power factor of superionic fcc H\textsubscript{2}O surpasses that of many conventional ionic/electronic thermoelectric materials (Supplementary Fig. S29)\cite{main038,main039}, suggesting that superionic fcc H\textsubscript{2}O may constitute an efficient mechanism for energy conversion and charge redistribution in planetary interiors.

\begin{figure*}[tp]
\centering
\includegraphics[width=0.85\textwidth]{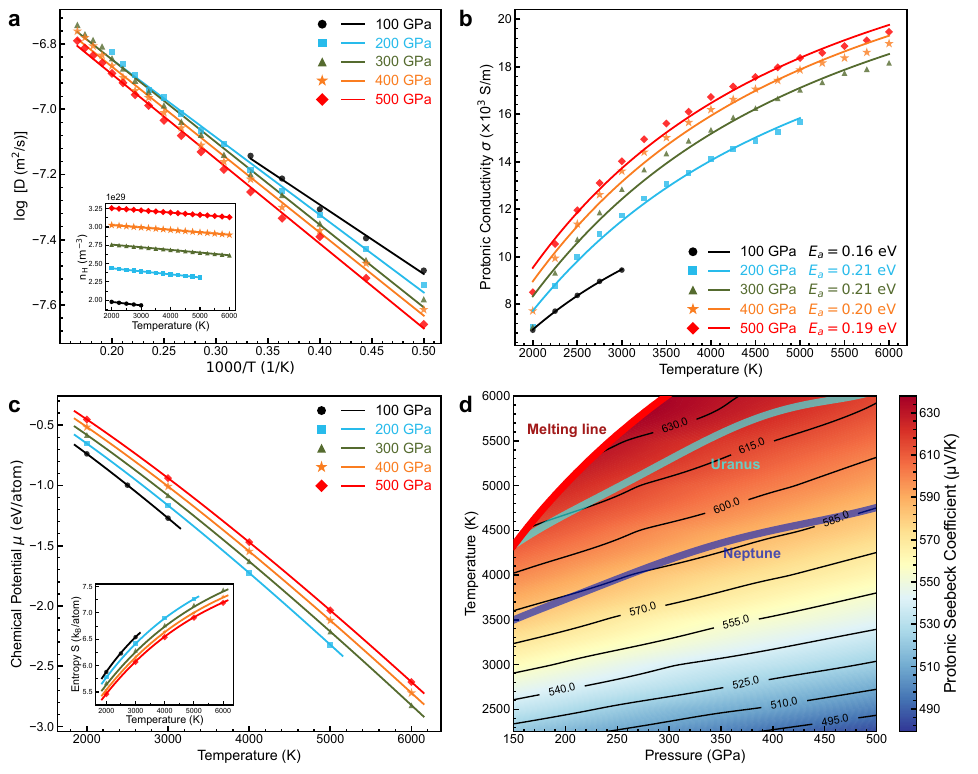}
\caption{\textbf{Protonic thermoelectric and transport properties of superionic fcc H\textsubscript{2}O ice under the conditions relevant to the interiors of Uranus and Neptune.} The colors of the curves and symbols correspond to temperature along isobars. The solid curves and symbols for the 100--500 GPa range are represented by black, blue, olive green, orange, and red, respectively. Also, circles, squares, triangles, stars, and diamonds are used to represent scatters for the 100--500 GPa, respectively. \textbf{a,} The diffusion coefficient of proton was plotted as an Arrhenius equation with the relationship of log (\emph{D}\textsubscript{H}) vs. 1000/\emph{T}. The inset plot shows the number density of protons as a function of temperature at 100--500 GPa. \textbf{b,} Protonic conductivity of superionic fcc H\textsubscript{2}O with increasing temperature at 100--500 GPa. The \emph{E}\textsubscript{a} is the activation enthalpy, which exhibits negligible pressure dependence. \textbf{c,} Temperature dependence of chemical potential and the entropy (inset) for proton at 100--500 GPa. \textbf{d,} Pressure and temperature contour plot of the protonic Seebeck coefficient with the averaged planetary thermal profiles of Uranus (U1 and U2) and Neptune (N1 and N2b) from Ref.~\citenum{main037} and the melting line.}
\label{fig:main3}
\end{figure*}

\subsection*{Magnetic Field Generation in Uranus and Neptune}

The highly eccentric and strongly non-dipolar magnetic field morphologies of Uranus and Neptune are thought to arise from a thin, active convective shell in their shallow interiors, underlain by a stable inner layer. A key implicit premise of this model is that the thin shell, composed of ionic fluid H\textsubscript{2}O, must simultaneously possess sufficiently high electrical conductivity and intense turbulent convection to sustain an effective global dynamo process. Computational and shock compression results suggest the conductivity of fluid H\textsubscript{2}O in the interiors of icy giants exceeds \ensuremath{\sim}1000 S/m\cite{main040}. Such conductivity values are sufficient to produce Reynolds numbers greater than 40 and sustain dynamo action in these regions\cite{main041,main042}. However, recent static conductivity measurements of dense ionic fluid H\textsubscript{2}O at 11--45 GPa and 1030--2750 K indicate that the ionic fluid H\textsubscript{2}O in the shallow thin layer exhibits a conductivity of 10--40 S/m\cite{main041}, which is much lower than values measured from shock-compression at 28--59 GPa and \ensuremath{\sim}2500--5000 K along the Hugoniot\cite{main040}. At these elevated temperatures, H\textsubscript{2}O enters a partially electronically conducting regime that is not representative of adiabatic thermal profiles in the shallow ionic fluid layer in Uranus and Neptune (2000--3000 K under standard adiabatic thermal profiles\cite{main037,main043}). Accordingly, the electrical conductivity of the thin fluid layer may be 1--2 orders of magnitude lower than previously estimated, imposing more stringent conditions for sustaining the dynamo and continuously generating the magnetic field. Furthermore, constraints from the interior structure and thermal evolution suggest that Uranus and Neptune may contain significant compositional gradients and thermal boundary layer structures\cite{main044,main045,main046}, which might inhibit large-scale convection and result in a fainter radiated luminosity of Uranus\cite{main047}. On the other hand, numerical simulations also show that complex chemical mixtures (H\textsubscript{2}O--CH\textsubscript{4}--NH\textsubscript{3})\cite{main012,main013,main014} may form stable stratified regions under high pressure, resulting in lower thermal transport efficiency to meet the requirements of the thin-shell model\cite{main007,main048}.

Thermoelectricity describes the direct conversion of heat into electricity without mechanical motion. This mechanism may become particularly significant for planets where low heat flux is insufficient to drive the convection of conductive fluids. Furthermore, it has also been applied to understanding the generation of the geomagnetic field\cite{main049}. Here, we found that superionic fcc H\textsubscript{2}O, which occupies the icy mantles above the rocky cores of Uranus and Neptune (Figs. 4a and 4b), presents a significant \emph{p}-TE effect. Therefore, proton migration can be driven by the temperature gradient inside the ice giants. Based on proton migration induced by the radial temperature gradient across the superionic ice layer, we utilized the Biot--Savart--Laplace law to calculate the planetary surface magnetic field strength. Our results show that the estimated magnetic field strength increases with the \emph{p}-TE conversion efficiency (\emph{\ensuremath{\eta}}). The theoretical maximum conversion efficiency (\emph{\ensuremath{\eta}}\textsubscript{max}) is determined by the thermoelectric figure of merit \emph{zT} (Supplementary Figs. S30 and S31), a dimensionless quantity that characterizes the capacity of a material to convert heat into electricity, and the temperature difference across the icy mantle. Thus, the predicted \emph{\ensuremath{\eta}}\textsubscript{max} can reach approximately 10\% under the thermodynamic conditions relevant to Uranus and Neptune. By combining calculations of protonic conductivity along the internal thermal profiles of Uranus and Neptune\cite{main037}, we fitted and annotated the contours of their internal protonic conductivity (\emph{\ensuremath{\sigma}}) and temperature gradient (\ensuremath{\nabla}\emph{\textsubscript{r}T}) relationship in the diagram (Supplementary Figs. S32 and S33) with different \emph{\ensuremath{\eta}} of 1\%, 2.5\%, 5\%, and 10\%. The predicted maximum field strength (\emph{\ensuremath{\eta}=}10 \%) is 80-100 \textmu T for Uranus and 60-80 \textmu T for Neptune, which is much stronger than the observations. However, thermal inhomogeneities, lateral temperature gradients, and compositional complexity may reduce the effective conversion efficiency below its ideal theoretical maximum in realistic planetary interiors. Our calculations indicate that a conversion efficiency of 2.5\% (Fig. 4c) is sufficient to produce a magnetic field strength comparable to available observations. The efficiency \emph{\ensuremath{\eta}} = 2.5\% is adopted here as an illustrative value. It is physically reasonable for superionic H\textsubscript{2}O, lying well below the theoretical maximum of approximately 10\% obtained from \emph{zT} for the relevant temperature differences, and it reproduces the observed field magnitudes. The comparison with the Voyager 2 field strengths therefore demonstrates that the \emph{p}-TE mechanism can account for the observed field magnitudes for physically plausible efficiencies. The results indicate that within the estimated range for Uranus's interior, the corresponding magnetic field strength values primarily fall near the 20--25 \textmu T contour lines, which are in good agreement with the Uranus's averaged surface magnetic field strength (23 \textmu T)\cite{main002} inverted by Voyager 2. On the other hand, the calculated magnetic field strength based on the \emph{\ensuremath{\sigma}}-\ensuremath{\nabla}\emph{\textsubscript{r}T} relation for Neptune is distributed between 15--20 \textmu T and is also consistent with the observed average magnetic field strength of Neptune (13 \textmu T)\cite{main003}. It is noteworthy that differences in the thermodynamic conditions of Uranus and Neptune produce distinct internal temperature gradients and electrical conductivity distributions, which in turn lead to the different magnetic field strengths, reflecting that Uranus can obtain a slightly higher average surface field strength than Neptune. This trend not only aligns with observations but also demonstrates that the thermoelectric circuit can naturally distinguish the magnetic field strengths of Uranus and Neptune.

The \emph{p}-TE mechanism provides a perspective on the origin of the magnetic fields in ice giants. This mechanism can be driven solely by internal temperature gradients, which does not require vigorous convective motions.~In contrast, the thin-shell dynamo model\cite{main008} relies on turbulent convection in the outer ionic fluid layer to generate the observed multipolar magnetic field. Although this model successfully reproduces the non-dipolar, non-axisymmetric field morphology, it requires both a sufficiently high electrical conductivity and a pre-existing initial magnetic field to initiate dynamo action---conditions that are difficult to satisfy simultaneously in the ionic fluid layer. Therefore, the \emph{p}-TE mechanism operates in the deeper superionic ice layer, where the radial temperature gradient drives proton migration and the generated thermoelectric currents can produce the magnetic field. This initial field, together with the mean dipolar field contribution from the \emph{p}-TE effect, might then be injected into the overlying ionic fluid layer, where the thin-shell dynamo may amplify and reorganize it into the observed multipolar structure. In this case, the interaction between these two mechanisms may be critical in understanding the complex magnetic fields of ice giants. The two have not yet been shown to be consistent within one coupled model, and a coupled treatment would require planetary-scale dynamo simulations.

Although \emph{p}-TE can generate significant magnetic field strengths comparable to the observations, our calculation only considers the thermoelectric currents generated by fcc superionic H\textsubscript{2}O, which is one component of the complex composition of ice giants. Furthermore, recent experiments have revealed that superionic H\textsubscript{2}O can adopt hexagonal close-packed (hcp) and mixed fcc/hcp stacking structures\cite{main050,main051}, adding further structural complexity to the icy mantles. Therefore, stacking sequence differences may give rise to quantitative deviations in protonic conductivity, thermal conductivity, and the Seebeck coefficient. Moreover, the hcp and mixed fcc/hcp stacking structures might lead to anisotropy in the transport properties. These variations could introduce additional spatial inhomogeneities in the thermoelectric current distribution, potentially affecting both the strength and the multipolar structure of the generated magnetic field. Moreover, the composition of the icy mantles might be mixed with complex ices such as NH\textsubscript{3} and CH\textsubscript{4}\cite{main012,main013,main014}. Recent studies also suggest possible reactions between H\textsubscript{2}O and rocks to generate superionic hydrous phases\cite{main052}. This complex compositional heterogeneity leads to variations in local electrical and thermal conductivity, which in turn may cause variations of Seebeck coefficients in the icy mantles. In addition, lateral variations in temperature of the icy mantles may also lead to spatial fluctuations in the temperature gradient. The inhomogeneity caused by compositional differences and variations in temperature gradients in the icy mantles may affect the 3-dimensional spatial distribution of both the intensity and direction of the \emph{p}-TE current, resulting in spatially complex magnetic fields of Uranus and Neptune. Therefore, the model is intended to account for the magnitude of the mean field and for its difference between the two planets, and not for the non-dipolar geometry of the observed fields. In the future, exploring the interplay between \emph{p-TE} currents in the icy mantle and the upper conductive fluid layer, and the prospective generation mechanism of multipolar magnetic fields, may provide pivotal insights on the origin of strange magnetic fields inside Uranus and Neptune.

\begin{figure*}[tp]
\centering
\includegraphics[width=0.88\textwidth]{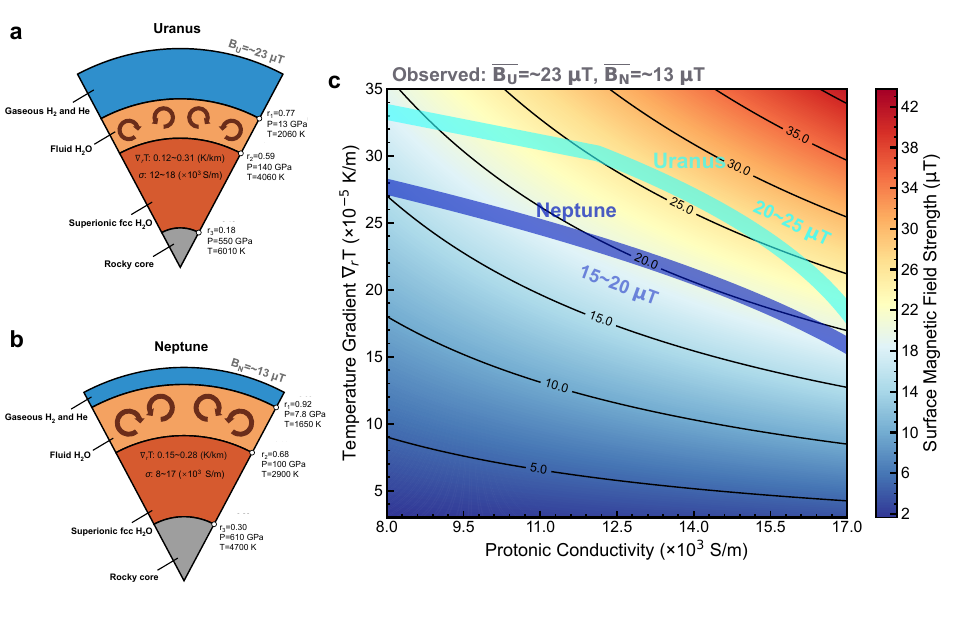}
\caption{\textbf{The thermodynamic profiles of ice giants and the dependence of surface magnetic field strength on protonic conductivity and temperature gradient. a,} Uranus' and \textbf{b,} Neptune's thermodynamic profiles of interior structures\cite{main043}. The overlying ionic fluid layer and the deeper superionic fcc H\textsubscript{2}O layer are indicated, illustrating the spatial coupling between the thermoelectric current generation in the superionic layer and the thin-shell dynamo action in the convective layer above. \textbf{c,} The surface magnetic strength as a function of protonic conductivity (\emph{\ensuremath{\sigma}}) and radial temperature gradient (\ensuremath{\nabla}\emph{\textsubscript{r}T}) for the thermoelectric conversion efficiency of \emph{\ensuremath{\eta}=}2.5\%. The cyan and dark blue shaded bands highlight the predicted surface magnetic field strength based on the protonic conductivity and temperature gradient parameters of Uranus and Neptune\cite{main037}. These regions present great consistence with the observed surface magnetic field strength of Uranus (\({\overline{B}}_{U}\)\ensuremath{\approx}23 \textmu T)\cite{main002} and Neptune (\({\overline{B}}_{N}\)\ensuremath{\approx}13 \textmu T)\cite{main003}.}
\label{fig:main4}
\end{figure*}

\section*{Methods}

\subsection*{\emph{Ab initio} calculations on the H\textsubscript{2}O phase}

All~\emph{ab initio}~calculations were performed within the framework of Kohn-Sham Density Functional Theory (KS-DFT)\cite{main053,main054}. The calculations were executed with the open-source software package ABACUS v3.8.4\cite{main055,main056,main057}, in which the exchange-correlation energy was described by the Perdew-Burke-Ernzerhof (PBE) functional within the generalized gradient approximation (GGA)\cite{main058}, augmented by Grimme's D3 dispersion correction\cite{main059}. The electronic states were represented by SG15 Optimized Norm-Conserving Vanderbilt (SG15-ONCV) pseudopotentials\cite{main060}, which treat the 1\emph{s}\textsuperscript{2}2\emph{s}\textsuperscript{2}2\emph{p}\textsuperscript{4} and 1\emph{s}\textsuperscript{1} configurations as valence electrons for O and H, respectively, in combination with a Linear Combination of Atomic Orbitals (LCAO) basis set of double-zeta plus polarization (DZP) numerical atomic orbitals\cite{main061}.

We employed periodic simulation cells with 54 and 108 water molecules for bcc and fcc phases, respectively. For all calculations, the kinetic energy cutoff and K-spacing were set to 100 Ry and 0.2 Bohr\textsuperscript{-1}, respectively, while the self-consistent field (SCF) convergence threshold was 10\textsuperscript{-6} Ry in total electronic energy. The~\emph{ab initio}~molecular dynamics (AIMD) simulations were conducted with the NVT ensemble, in which the temperature was controlled using the Langevin thermostat implemented by ABACUS. To reproduce the targeted thermodynamic regime, lattice parameters were adjusted so that trajectories could be generated across 100--500 GPa and 1000--8000 K conditions at a timestep of 0.5 fs.

\subsection*{DP neural network model}

The interatomic potential was constructed within the DP framework by employing the descriptor ``se\_e2\_a''\cite{main062,main063}, as implemented in the DeePMD-kit code package (v2.2.8)\cite{main064,main065,main066}. The initial training datasets for the DP model were first generated by randomly sampling snapshots from the AIMD trajectories. The training datasets were then extended by using an active-learning workflow of the DP-Generator (DP-GEN) package\cite{main067}, in which candidate configurations were iteratively explored and collected. Under this workflow, each iteration consisted of three steps: exploration, labeling, and training.

In order to comprehensively explore configurations covering the thermodynamic conditions in the interiors of Uranus and Neptune, the exploration stage was performed using NPT simulations spanning 70--600 GPa and 1000--8000 K. In the exploration stage, configurations were selected for labeling through pre-set lower and upper trust levels for force deviation, in which the lower and upper trust levels were set to 0.3 and 0.8 eV/Å, respectively. After the exploration step of each iteration, the farthest-point sampling (FPS)\cite{main068} algorithm was employed to select representative subsets from the candidate configurations for DFT calculations using ABACUS. For the training step of each iteration, the neural network framework was comprised of embedding layers of sizes \{25, 50, 100\} and fitting layers of sizes \{240, 240, 240\}. A cutoff radius of 6 Å, which was smoothly decayed to 0.5 Å with a 1/\emph{r} function, was employed to describe the local atomic environment. Finally, a total of 32828 frames of both bcc and fcc H\textsubscript{2}O were collected from all DP-GEN iterations. From these, a total of 2127 frames were randomly selected from the bcc and fcc datasets as the test set, with the remaining data used as the training set for the final DP model.

\subsection*{DPMD simulations}

All DPMD calculations of bcc and fcc H\textsubscript{2}O were performed by using the Large-scale Atomic/Molecular Massively Parallel Simulator (LAMMPS) package\cite{main069,main070,main071}, which was interfaced with the DP module. Throughout the simulations, the system pressure and temperature were controlled by the Langevin thermostat and Nosé-Hoover barostat with the damping parameters of 50 and 500 fs, respectively\cite{main072}. Additionally, a 0.5 fs time step was selected for integration of the equations of motion to preserve the stability of the atomic trajectories over the simulations. The simulation cell volumes were determined by performing 20 ps DPMD simulations in the NPT ensemble at the target temperature and pressure, with the average box size taken from the second half of the simulation trajectory.

\subsection*{Gibbs free energy calculations}

The Gibbs free energies of the bcc, fcc, and liquid H\textsubscript{2}O phases were computed using the non-equilibrium thermodynamic integration (NETI) method\cite{main023,main073,main074}. The solid bcc H\textsubscript{2}O phase, containing 3456 oxygen ions and 6912 protons, was calculated using the Einstein crystal as a reference system with an initial thermodynamic reference point of 300 GPa and 1000 K. The superionic fcc H\textsubscript{2}O phase, containing 4000 oxygen ions and 8,000 protons, was calculated using the Einstein crystal and Uhlenbeck--Ford (UF) fluids as reference systems for the oxygen ions and protons, respectively\cite{main075,main076,main077}, with an initial thermodynamic reference point of 300 GPa and 5000 K. The fcc-liquid H\textsubscript{2}O phase was calculated using the UF fluid as a reference system with an initial thermodynamic reference point of 300 GPa and 7500 K.

For the Gibbs free energy calculations of each phase, the equilibrium time~\emph{t}\textsubscript{eq}~was set to 25 ps to ensure system equilibrium, and the switching time~\emph{t}\textsubscript{sw}, during which the system is transformed from the reference system to the DP model via the coupling parameter~\emph{\ensuremath{\lambda}} (0\textless{}\emph{\ensuremath{\lambda}\textless1}), was set to 100 ps. The final Gibbs free energy was determined from 10 independent NETI runs.

Finally, the phase boundaries between the bcc, fcc, and liquid H\textsubscript{2}O phases were determined by starting from the two-phase coexistence point, calculating the differences of enthalpy and volume at each pressure and temperature point, and applying the Clausius--Clapeyron equation (CCE) to calculate the coexistence lines (see the Supplementary Information for details).

\subsection*{Non-equilibrium molecular dynamics}

The lattice thermal conductivity of the bcc and fcc phases was computed using non-equilibrium molecular dynamics (NEMD) based on the Müller-Plathe (MP) method\cite{main078,main079}, as implemented by the DP model. According to Fourier's law of heat conduction, the thermal conductivity~\emph{\ensuremath{\kappa}}~is expressed as:

\begin{equation}
\kappa = - \frac{\langle J(t)\rangle}{\langle\partial T/\partial z\rangle},
\label{eq:main01}
\end{equation}

where the time average of the heat flux is denoted as \(\langle J(t)\rangle\), and the time average of the temperature gradient in the simulation cell is represented as \(\langle\partial T/\partial z\rangle\).

In the MP framework, the heat flux is generated by the periodic exchange of atomic velocities between two predefined regions within the simulation cell, designated as the ``hot region'' and the ``cold region.'' At a fixed interval, the hottest atom in the ``cold region'' and the coldest atom in the ``hot region'' are selected, and their velocities are exchanged according to the following relations:

\begin{equation}
\mathbf{v}_{\mathrm{cold}}^{\mathrm{new}} = - \mathbf{v}_{\mathrm{cold}} + 2\left\lbrack \frac{m_{\mathrm{cold}}\mathbf{v}_{\mathrm{cold}} + m_{\mathrm{hot}}\mathbf{v}_{\mathrm{hot}}}{m_{\mathrm{cold}} + m_{\mathrm{hot}}} \right\rbrack,
\label{eq:main02}
\end{equation}

\begin{equation}
\mathbf{v}_{\mathrm{hot}}^{\mathrm{new}} = - \mathbf{v}_{\mathrm{hot}} + 2\left\lbrack \frac{m_{\mathrm{cold}}\mathbf{v}_{\mathrm{cold}} + m_{\mathrm{hot}}\mathbf{v}_{\mathrm{hot}}}{m_{\mathrm{cold}} + m_{\mathrm{hot}}} \right\rbrack,
\label{eq:main03}
\end{equation}

in which the atomic masses \emph{m}\textsubscript{cold} and \emph{m}\textsubscript{hot}, along with their initial velocities \textbf{v}\textsubscript{cold} and \textbf{v}\textsubscript{hot}, are used in the energy exchange, which induces a controlled heat flux while ensuring that the total energy of the system is conserved. Therefore, the lattice thermal conductivity expression in \textbf{Eq. (1)} will be modified as follows:

\begin{equation}
\kappa = - \frac{\sum_{\mathrm{transfers}}^{}{\sum_{\mu}^{}\frac{1}{2}}m_{\mu}\left( v_{\mu,\mathrm{hot}}^{2} - v_{\mu,\mathrm{cold}}^{2} \right)}{2tL_{x}L_{y}\langle\partial T/\partial z\rangle},
\label{eq:main04}
\end{equation}

where \emph{m\textsubscript{\ensuremath{\mu}}} is the mass of a particle of type \emph{\ensuremath{\mu}}, \emph{v\textsubscript{\ensuremath{\mu},}}\textsubscript{hot} and \emph{v\textsubscript{\ensuremath{\mu},}}\textsubscript{cold} are the velocities of the selected particle of type \emph{\ensuremath{\mu}} in the hot and cold regions, respectively, \emph{t} is the total effective duration over the NEMD simulation, and \emph{L\textsubscript{x}} and \emph{L\textsubscript{y}} are the lengths of the simulation cell along the \emph{x} and \emph{y} directions.

Due to boundary scattering caused by finite-size effects, a series of simulations were performed with different lengths \emph{L\textsubscript{z}} along the heat flux direction, while maintaining constant cross-sectional \emph{L\textsubscript{x}} and \emph{L\textsubscript{y}} sizes. The extrapolation to the infinite size was conducted by fitting the inverse thermal conductivity to the inverse length\cite{main080,main081}:

\begin{equation}
\kappa_{L}^{- 1} = \kappa_{\infty}^{- 1} + bL^{- 1},
\label{eq:main05}
\end{equation}

where~\emph{\ensuremath{\kappa}\textsubscript{L}}~is the lattice thermal conductivity for the cell length~\emph{L}~along the heat flux direction, and~\emph{\ensuremath{\kappa}}\textsubscript{\ensuremath{\infty}}~is the thermal conductivity in the infinite-length limit.

All the NEMD simulations were performed using the LAMMPS package. The initial configurations were equilibrated within the NVT ensemble for 200 ps. Subsequently, the simulation system was switched to the NVE ensemble, and a steady heat flux was imposed by exchanging momentum between the hot and cold regions at a momentum exchange interval of 25 fs. The simulation was conducted with a time step of 0.5 fs. The system was allowed to reach steady-state thermal transport over 300 ps, and the following 1.0 ns simulation was used to collect the temperature gradient.

\subsection*{Diffusion coefficient and protonic conductivity}

The protonic transport properties were investigated by employing DPMD simulations, in which the mean square displacements (MSDs) method was used to calculate the diffusion coefficients of the species. The simulations were executed in the NVT ensemble. The MSDs of each ionic species can be computed as:

\begin{equation}
\langle\left\lbrack \overrightarrow{\mathbf{r}}(t) \right\rbrack^{2}\rangle = \frac{1}{N}\sum_{i = 1}^{N}\langle\left| {\overrightarrow{\mathbf{r}}}_{i}(t) - {\overrightarrow{\mathbf{r}}}_{i}(0) \right|^{2}\rangle,
\label{eq:main06}
\end{equation}

where \emph{N} is the number of target ions, and \({\overrightarrow{\mathbf{r}}}_{i}(t)\) represents the position of the \emph{i}-th ion at time \emph{t}. The diffusion coefficient \emph{D} was then obtained from the long-time linear regime of the MSDs according to the Einstein relation:

\begin{equation}
D = \frac{1}{6}\frac{d}{dt}\left\langle \left\lbrack \overrightarrow{\mathbf{r}}(t) \right\rbrack^{2} \right\rangle,
\label{eq:main07}
\end{equation}

The diffusion coefficients calculated at various temperatures were fitted with an Arrhenius equation:

\begin{equation}
D(T) = D_{0}\exp\left( - \frac{\Delta H}{k_{\mathrm{B}}T} \right),
\label{eq:main08}
\end{equation}

where \ensuremath{\Delta}\emph{H} is the activation enthalpy, \emph{D}\textsubscript{0} is a pre-exponential factor, \emph{k}\textsubscript{B} is the Boltzmann constant, and \emph{T} is the temperature.

Based on the diffusion coefficients, the ionic conductivity \emph{\ensuremath{\sigma}} was estimated via the Nernst-Einstein equation,

\begin{equation}
\sigma_{i}(T) = \frac{fn_{i}q_{i}^{2}}{k_{\mathrm{B}}T}D_{i}(T),
\label{eq:main09}
\end{equation}

in which~\emph{f}~is a numerical factor approximately equal to unity,~\emph{n\textsubscript{i}}\textsubscript{~}denotes the number density of transport ions, and~\emph{q\textsubscript{i}}\textsubscript{}~is their charge.

\subsection*{Protonic thermodiffusion and thermopower}

All calculations of ionic thermoelectric effects in the superionic fcc phase are performed within the framework of Onsager relations\cite{main030,main082,main083}. In the electrolyte system, the ionic flux is finally determined as:

\begin{equation}
J_{i,z} = - D_{i}\left( \frac{\partial n_{i}}{\partial z} + \frac{q_{i}n_{i}}{k_{\mathrm{B}}T}E_{z} + \frac{n_{i}}{k_{\mathrm{B}}T}\frac{\partial\mu_{i}}{\partial z} \right),
\label{eq:main10}
\end{equation}

where \emph{E\textsubscript{z}} represents the hypothetical thermoelectric field and \emph{\ensuremath{\mu}\textsubscript{i}} represents the chemical potential of \emph{i}-species. The relationship between the chemical potential of the \emph{i}-species and the temperature gradient is expressed as follows\cite{main084,main085}:

\begin{equation}
\frac{\partial\mu_{i}}{\partial z} = \frac{\partial\mu_{i}}{\partial T}\frac{\partial T}{\partial z} = {\widehat{S}}_{i}\frac{\partial T}{\partial z},
\label{eq:main11}
\end{equation}

where~\({\widehat{S}}_{i} = \partial\mu_{i}/\partial z\)~is the Eastman transfer entropy\cite{main086} of the~\emph{i}-species, derived from the temperature dependence of the chemical potential.

At steady state, the current in the circuit is zero, then:

\begin{equation}
\sum_{i}^{}q_{i}J_{i,z} = 0.
\label{eq:main12}
\end{equation}

\begin{widetext}
\begin{equation}
- \sum_{i}^{}D_{i} \cdot \left( q_{i}\frac{\partial n_{i}}{\partial z} + \frac{q_{i}^{2}n_{i}}{k_{\mathrm{B}}T}\frac{\partial V}{\partial z} + \frac{q_{i}{\widehat{S}}_{i}n_{i}}{k_{\mathrm{B}}T}\frac{\partial T}{\partial z} \right) = 0.
\label{eq:main13}
\end{equation}
\end{widetext}

Generally, the ion concentration gradient is assumed to be negligible. Thus, \textbf{Eq. (10)} simplifies to:

\begin{equation}
\sum_{i}^{}q_{i}D_{i}\left( \frac{q_{i}n_{i}}{k_{\mathrm{B}}T}E_{z} + \frac{n_{i}{\widehat{S}}_{i}}{k_{\mathrm{B}}T}\frac{\partial T}{\partial z} \right) = 0.
\label{eq:main14}
\end{equation}

Solving for \emph{E\textsubscript{z}} yields:

\begin{equation}
E_{z} = - \frac{\sum_{i}^{}q_{i}n_{i}{\widehat{S}}_{i}D_{i}}{\sum_{i}^{}q_{i}^{2}n_{i}D_{i}}\frac{\partial T}{\partial z}.
\label{eq:main15}
\end{equation}

In the superionic fcc phase, oxygen atoms occupy the sublattice sites with negligible diffusion (\emph{D}\textsubscript{O}\ensuremath{\approx}0), so that the numerator of \textbf{Eq. (15)} contains only the hydrogen contribution. Using the relation \(E_{z} = - \partial V/\partial z\), the induced potential gradient is obtained as:

\begin{equation}
\frac{\partial V}{\partial z} = - \frac{{\widehat{S}}_{\mathrm{H}}}{q_{\mathrm{H}}}\frac{\partial T}{\partial z}.
\label{eq:main16}
\end{equation}

Thus, the protonic thermoelectric (Seebeck) coefficient of superionic fcc H\textsubscript{2}O is given by:

\begin{equation}
S = - \frac{\partial V/\partial z}{\partial T/\partial z} = \frac{{\widehat{S}}_{\mathrm{H}}}{q_{\mathrm{H}}}
\label{eq:main17}
\end{equation}

The Seebeck coefficient is conventionally defined with a negative sign, such that a positive temperature gradient induces an opposing thermoelectric field at steady state. The chemical potential is calculated by employing the Widom insertion method\cite{main087},

\begin{widetext}
\begin{equation}
\mu_{i} = k_{\mathrm{B}}T\ln\left( \rho_{i}{(\frac{h^{2}}{2\pi m_{i}k_{\mathrm{B}}T})}^{\frac{3}{2}} \right) - k_{\mathrm{B}}T\ln\left( \left. \  < \exp\left( - \frac{U_{N_{i} + 1} - U_{N_{i}}}{k_{\mathrm{B}}T} \right) \right.\  > \right)
\label{eq:main18}
\end{equation}
\end{widetext}

where \emph{\ensuremath{\rho}\textsubscript{i}} is the number density of the \emph{i}-species, \(U_{N_{i}}\) and \(U_{N_{i} + 1}\) are the potential energies of the system with \emph{N} and \emph{N}+1 \emph{i}-species particles, respectively.

\subsection*{Magnetic field for icy mantles}

To evaluate the magnetic field generated by thermoelectrically driven currents in the superionic ice shell, we established a mathematical framework in spherical coordinates based on the following simplifying assumptions: (i) the temperature gradient is purely radial and spatially uniform across the shell; (ii) the system is in steady state, and the charge density of the sphere model does not vary with time; (iii) the transport coefficients (thermopower \emph{S}, protonic conductivity \emph{\ensuremath{\sigma}}, and thermal conductivity \emph{\ensuremath{\kappa}}) are affected by the radial temperature within the superionic layer and vary along the radial direction. These assumptions reduce the problem to a one-dimensional (1D) radial thermoelectric current model, which represents the leading order mean dipolar field contribution.

The superionic ice layer was treated as a spherical shell with an outer radius \emph{R}\textsubscript{SI}, across which a temperature gradient was assumed to exist in the radial direction. Therefore, the charge-continuity equation in spherical coordinates can be written as:

\begin{equation}
\nabla \cdot \overrightarrow{\mathbf{J}} + \frac{\partial\rho_{e}}{\partial t} = 0.
\label{eq:main19}
\end{equation}

where the \(\rho_{e}\) is the charge density and \(\overrightarrow{\mathbf{J}}\) is the current density vector. On a planetary scale, it is assumed that the charge within the superionic ice layer does not change with time, i.e.:

\begin{equation}
\frac{\partial\rho_{e}}{\partial t} \approx 0.
\label{eq:main20}
\end{equation}

Finally, the current density vector in spherical coordinates can be expressed as:

\begin{widetext}
\begin{equation}
\frac{1}{r^{2}}\frac{\partial}{\partial r}\left( r^{2}J_{r} \right) + \frac{1}{r\sin\theta}\frac{\partial}{\partial\theta}\left( \sin\theta J_{\theta} \right) + \frac{1}{r\sin\theta}\frac{\partial J_{\phi}}{\partial\phi} = 0.
\label{eq:main21}
\end{equation}
\end{widetext}

To simplify the derivation, we assumed that the current density is axisymmetric along the \emph{\ensuremath{\phi}} direction, i.e., \ensuremath{\partial}\emph{J\textsubscript{\ensuremath{\phi}}}/\ensuremath{\partial}\emph{\ensuremath{\phi}} = 0. Under this condition, the \emph{J\textsubscript{\ensuremath{\theta}}} component is decoupled from the governing equations and can therefore be expressed solely as a function of \emph{r} and \emph{\ensuremath{\theta}}, i.e., \emph{J\textsubscript{\ensuremath{\theta}}} = \emph{J\textsubscript{\ensuremath{\theta} }}(\emph{r}, \emph{\ensuremath{\theta}}), which is independent of the \emph{\ensuremath{\phi}} component.

The azimuthal current component \emph{J\textsubscript{\ensuremath{\phi}}} is the primary contributor to the magnetic dipole moment. In a spherically symmetric thermoelectric model where the driving force (the radial temperature gradient) is uniform, the resulting radial current density \emph{J\textsubscript{r}} is also axisymmetric. For such a radial current at polar angle \emph{\ensuremath{\theta}} the equivalent azimuthal (toroidal) current density around the symmetry axis is obtained by projecting the radial direction onto the equatorial plane\cite{main049}.

\begin{equation}
J_{\phi} = J_{\phi}(r,\ \theta) \approx J_{r}\sin\theta.
\label{eq:main22}
\end{equation}

This approximation describes the azimuthal current induced by the deflection of charge carriers along an isothermal spherical shell. The radial current itself is determined by the thermoelectric effect:

\begin{equation}
J_{r} = \sigma S\frac{\partial T}{\partial r},
\label{eq:main23}
\end{equation}

where \emph{S}, \emph{\ensuremath{\sigma}} and \(\partial T/\partial r\) are the thermopower, ionic conductivity, and radial temperature gradient, respectively.

At the planetary scale, the spherical shell current distribution can be approximated by an equivalent magnetic dipole moment. This dipole approximation captures the leading-order (dipolar) component of the surface magnetic field, which is the dominant contribution for the mean field amplitude. Therefore, the magnetic moment generated by the current density in the superionic ice layer can be expressed as:

\begin{equation}
\overrightarrow{\mathbf{m}} = \frac{1}{2}\int_{V}^{}{\overrightarrow{\mathbf{r}}}^{'} \times \overrightarrow{\mathbf{J}}\left( {\overrightarrow{\mathbf{r}}}^{'} \right)dV^{'}.
\label{eq:main24}
\end{equation}

Furthermore, it can be expanded as:

\begin{widetext}
\begin{equation}
m_{x} = - \frac{1}{2}\int_{r_{\mathrm{in}}}^{r_{\mathrm{out}}}{\int_{0}^{\pi}{\int_{0}^{2\pi}{{r^{'}}^{3}\left( J_{\theta}\sin\phi + J_{\phi}\left( r^{'},\theta \right)\cos\theta\cos\phi \right)}\sin\theta}}d\phi d\theta dr^{'},
\label{eq:main25}
\end{equation}
\end{widetext}

\begin{widetext}
\begin{equation}
m_{y} = - \frac{1}{2}\int_{r_{\mathrm{in}}}^{r_{\mathrm{out}}}{\int_{0}^{\pi}{\int_{0}^{2\pi}{{r^{'}}^{3}\left( - J_{\theta}\cos\phi + J_{\phi}\left( r^{'},\theta \right)\cos\theta\sin\phi \right)}\sin\theta}}d\phi d\theta dr^{'},
\label{eq:main26}
\end{equation}
\end{widetext}

\begin{widetext}
\begin{equation}
m_{z} = + \frac{1}{2}\int_{r_{\mathrm{in}}}^{r_{\mathrm{out}}}{\int_{0}^{\pi}{\int_{0}^{2\pi}{{r^{'}}^{3}J_{\phi}\left( r^{'},\theta \right)}\sin^{2}\theta}}d\phi d\theta dr^{'}.
\label{eq:main27}
\end{equation}
\end{widetext}

Considering spherical symmetry, the integrals of cos\emph{\ensuremath{\phi}} and sin\emph{\ensuremath{\phi}} over {[}0, 2\ensuremath{\pi}{]} vanish:

\begin{equation}
\int_{0}^{2\pi}\cos\phi d\phi = \int_{0}^{2\pi}{\sin\phi}d\phi = 0,
\label{eq:main28}
\end{equation}

leading to \emph{m\textsubscript{x}} = \emph{m\textsubscript{y}} = 0, and the \emph{z}-component can be approximated by evaluating the geometric factor \emph{r}\ensuremath{\prime}\textsuperscript{3} at the characteristic radius \emph{R}\textsubscript{SI} of the superionic ice shell and taking it outside the radial integral as:

\begin{widetext}
\begin{equation}
m_{z} \approx \frac{1}{2}R_{\mathrm{SI}}^{3}\int_{r_{\mathrm{in}}}^{r_{\mathrm{out}}}{\int_{0}^{\pi}{\int_{0}^{2\pi}{J_{\phi}\left( r^{'},\ \theta \right)}\sin^{2}\theta}}d\phi d\theta dr^{'}.
\label{eq:main29}
\end{equation}
\end{widetext}

where \emph{R}\textsubscript{SI} is the characteristic radius of the superionic ice shell.

Thus, according to the Biot--Savart law, the differential form of the surface magnetic field of a planet can be expressed as:

\begin{equation}
\text{d}\mathbf{B} = \frac{\mu_{0}}{4\pi}\frac{\overrightarrow{\mathbf{J}} \times \overrightarrow{\mathbf{R}}}{R^{3}}\text{d}V^{'}.
\label{eq:main30}
\end{equation}

Integrating both sides of \textbf{Eq. (30)} over the volume of the current-carrying region yields the total magnetic field at the observation point. Within the framework of the dipole approximation, the magnetic flux density can finally be calculated as:

\begin{equation}
\mathbf{B}\left( \overrightarrow{\mathbf{R}} \right) \approx \frac{\mu_{0}}{4\pi R^{3}}\left( 3\overrightarrow{\mathbf{e}_{r}}\left( \overrightarrow{\mathbf{m}} \cdot \overrightarrow{\mathbf{e}_{r}} \right) - \overrightarrow{\mathbf{m}} \right),
\label{eq:main31}
\end{equation}

where \(\mu_{0}\) is the permeability of free space and \emph{R} represents the planetary radius. Based on \textbf{Eq. (31)}, the components of the surface magnetic field strength can be further expressed as:

\begin{equation}
B_{r}(R,\theta) = \frac{\mu_{0}}{4\pi}\frac{2m_{z}\cos\theta}{R^{3}},
\label{eq:main32}
\end{equation}

\begin{equation}
\quad B_{\theta}(R,\theta) = \frac{\mu_{0}}{4\pi}\frac{m_{z}\sin\theta}{R^{3}}.
\label{eq:main33}
\end{equation}

Therefore, the surface magnetic field strength can be expressed as \(B = \mu_{0}\left| \overrightarrow{\mathbf{m}} \right|/(4\pi R^{3})\).

\subsection*{Efficiency evaluation of \emph{p}-TE and constraints on current density}

The local thermoelectric power density, defined as the electrical power generated per unit volume during the thermoelectric process, can be given by the dot product of the electric current and the electric field:

\begin{equation}
p_{\text{elec }}(r) = \overrightarrow{\mathbf{J}} \cdot \overrightarrow{\mathbf{E}}.
\label{eq:main34}
\end{equation}

In the 1D radial model considered here, \(\overrightarrow{\mathbf{J}} \cdot \overrightarrow{\mathbf{E}}\) reduces to the product of the radial components, \emph{J\textsubscript{r} E\textsubscript{r}}. The thermoelectric field generated by the Seebeck effect along the radial direction is \emph{E\textsubscript{r }}= \emph{S}\ensuremath{\partial}\emph{T} /\ensuremath{\partial}\emph{r}. Substituting the radial current density \emph{J\textsubscript{r }}= \emph{\ensuremath{\sigma}S}\ensuremath{\partial}\emph{T} /\ensuremath{\partial}\emph{r} from \textbf{Eq. (23)} yields the local power density \emph{p}\textsubscript{elec}=\emph{\ensuremath{\sigma}S}\textsuperscript{2}(\ensuremath{\partial}\emph{T} /\ensuremath{\partial}\emph{r})\textsuperscript{2}. It can be integrated along the shell thickness to approximate the electrical power output per unit area through a thickness \ensuremath{\Delta}\emph{R}:

\begin{equation}
P_{\text{elec }\text{/}\text{ area }} \approx p_{\text{elec }}\mathrm{\Delta}R = \sigma S^{2}(\nabla T)^{2}\mathrm{\Delta}R,
\label{eq:main35}
\end{equation}

where \ensuremath{\Delta}\emph{R=r}\textsubscript{out}\emph{-r}\textsubscript{in} is the thickness of the superionic ice shell.

The heat flux available for thermoelectric conversion is assumed to follow Fourier's law of purely conductive heat transport, with a uniform thermal conductivity \emph{\ensuremath{\kappa}} within the superionic ice layer:

\begin{equation}
\left| Q_{\text{heat }} \right| = \kappa|\nabla T|.
\label{eq:main36}
\end{equation}

The electrical power output per unit area cannot exceed the fraction of the heat flux that can be converted into electrical work, as constrained by the second law of thermodynamics. This energy conservation constraint is given by:

\begin{equation}
P_{\text{elec }\text{/}\text{ area }} \leq \eta_{\max}\left| Q_{\text{heat }} \right|
\label{eq:main37}
\end{equation}

where \emph{\ensuremath{\eta}}\textsubscript{max} is the thermoelectric conversion efficiency, which is determined by the material's thermoelectric figure of merit \emph{zT} and the temperature difference across the shell. It can be calculated by using the following formula\cite{main088,main089}:

\begin{equation}
zT = \frac{S^{2}\sigma T}{\kappa}
\label{eq:main38}
\end{equation}

\begin{equation}
\eta_{\max} = \frac{T_{\text{H}} - T_{\text{C}}}{T_{\text{H}}}\frac{\sqrt{\overline{zT} + 1} - 1}{\sqrt{\overline{zT} + 1} + T_{\text{C}}\text{/}T_{\text{H}}}
\label{eq:main39}
\end{equation}

in which the \emph{zT}, \emph{T}\textsubscript{H} and \emph{T}\textsubscript{C} are thermoelectric figure of merit, hot end temperature and cold end temperature, respectively. Based on the above relation, an inequality is obtained, which further constrains the maximum value of the current density.

\begin{equation}
J_{r}^{\max} \leq \frac{\eta_{\max}\kappa}{S\mathrm{\Delta}R}
\label{eq:main40}
\end{equation}

This inequality provides an upper bound on the radial thermoelectric current density that can be sustained by the available heat flux under the maximum conversion efficiency \emph{\ensuremath{\eta}}\textsubscript{max} thereby linking the material transport properties (\emph{\ensuremath{\kappa}}, \emph{S}) and the shell geometry (\ensuremath{\Delta}\emph{R}) to the current-generating capacity of the superionic ice layer.

\section*{Data availability}

All the data and machine--learning potential used in this study are stored at \url{https://xxx.xxx.xxx}. Source Data are provided with this paper. And data needed to evaluate the conclusions in the paper are present in the paper and the Supplementary Information.

\section*{Code availability}

For the software packages used in this study, ABACUS software is openly available at \url{https://github.com/deepmodeling/abacus-develop}; DeePMD‐kit is developed openly at \url{https://github.com/deepmodeling/deepmdkit}; LAMMPS is developed openly at \url{https://github.com/lammps/lammps}.

\section*{Acknowledgments}

Numerical computations were supported by the Open--Source Supercomputing Center of S-A-I.

\section*{Funding Statement}

The project is supported by National Natural Science Foundation of China, 42350002 (to Y.H.). We acknowledge the support of the National Key Research and Development Program of China, Grant No. 2024YFF0807500 (to Y.H.) and acknowledge the support from "0-1" Original Research Project by the Institute of Geochemistry, Chinese Academy of Sciences, DHSZZ2023-2 (to Y.H.). C.Z. acknowledges National Natural Science Foundation of China (42230311).

\section*{Author contributions}

Conceptualization: Y.H., Methodology: D.L., W.Z., Y.H., Y.S. Investigation: D.L., Y.H. Visualization: D.L., C.Z. Funding acquisition: Y.H. Supervision: Y.H. Writing---original draft: D.L., Y.H. Writing---review \& editing: D.L., W.Z., Y.H., X.G., C.Z., Y.S.

\section*{Competing interests}

The authors declare that they have no competing interests.

\end{document}